# Quantum dynamics at finite temperature: Time-dependent quantum Monte Carlo study.

Ivan P. Christov

Physics Department, Sofia University, 1164 Sofia, Bulgaria

Email: ivan.christov@phys.uni-sofia.bg

**Abstract**

In this work we investigate the ground state and the dissipative quantum dynamics of interacting charged particles in an external potential at finite temperature. The recently devised time-dependent quantum Monte Carlo (TDQMC) method allows a self-consistent treatment of the system of particles together with bath oscillators first for imaginary-time propagation of Schrödinger type of equations where both the system and the bath converge to their finite temperature ground state, and next for real time calculation where the dissipative dynamics is demonstrated. In that context the application of TDQMC appears as promising alternative to the path-integral related techniques where the real time propagation can be a challenge.

## 1. Introduction

The area of non-equilibrium quantum statistical physics is still an object of rapid expansion in view of the new challenges posed by the recent developments in quantum transport in nano-materials, strongly correlated systems, dynamics of Bose-Einstein condensates at finite temperature, etc. The many-body open quantum systems have been treated perturbatively using the Green function formalism [1] and numerically by different Monte Carlo techniques [2,3]. Unlike in classical mechanics the incorporation of finite temperature effects is not trivial in quantum physics. The extension of the theoretical models at nonzero temperature involves environmental effects in the dynamics of the system by introducing nonlinearity into the Schrödinger equation itself [4] as well as using stochastic system-bath approaches where the quantum system which consists of relatively few degrees of freedom is coupled to a much larger bath through a system-bath coupling thus introducing quantum decoherence and dissipation [5]. The advantage of the latter is that it reduces in the classical limit to the familiar description based on generalized Langevin equation. However, most of the existing methods at finite temperature are focused on finding the ground state of the quantum system under consideration. Rare examples of real-time simulation techniques at finite temperature include multi configuration time-dependent self-consistent field [6] and real-time path integral [7]. Although time-polynomial the efficiency of the former method is hampered by the calculation of large number of Coulomb and exchange integrals while the latter may suffer from the rapid oscillations of the quantum propagator [8]. Therefore further development of non-perturbative time-dependent methods at finite temperature is indispensable to meet the challenges triggered by the experimental progress. A recent method with linear - to - low order polynomial scaling, the time-dependent quantum Monte-Carlo (TDQMC), combines self-consistently the non-stationary statistics of waves and particles (walkers) which evolve in physical space-time [9-11]. While the walker's distribution in space corresponds to the electron density, the quantum waves which guide those particles evolve according to a set of coupled nonlinear time-dependent Schrödinger equations. During the preparation stage of imaginary-time propagation TDQMC has a variational structure with the only variational parameter being the nonlocal quantum correlation length. Besides its conceptual simplicity TDQMC is unitary and numerically stable against rapid oscillations in the phase of the quantum state.

In this paper, the TDQMC method is applied to study finite-temperature effects in model quantum systems based on system-bath approach where a small system is coupled to a large bath of quantum oscillators at fixed temperature.

## 2. Preliminaries

For a non-relativistic system consisting of N particles of different kinds (quantum species) the Schrödinger equation reads:

$$i\hbar \frac{\partial}{\partial t} \Psi(\mathbf{R},t) = H \Psi(\mathbf{R},t), \tag{1}$$

where:

$$H = -\sum_i^N \frac{\hbar^2}{2M_i} \nabla_i^2 + V(\mathbf{R},t), \tag{2}$$

is the many-body quantum Hamiltonian, and $\mathbf{R} = \{\mathbf{r}_1,...,\mathbf{r}_N\}$ are the degrees of freedom. The potential in Eq. (2) may include electron-nuclear, nuclear-electron, electron-electron, nuclear-nuclear, and external potentials and $M_i$ is the mass of i-th species. Since, in general, the Hamiltonian in Eq. (2) is not separable in coordinates the numerical solution of Eq. (1) scales exponentially with the number of particles involved. Then, a factorization of the wave function of the type:

$$\Psi(\mathbf{r}_1, \mathbf{r}_2,...,\mathbf{r}_N, t) = \prod_{i=1}^N \varphi_i(\mathbf{r}_i, t) \tag{3}$$

reduces Eqs.(1),(2) to the well-known set of self-consistent mean field (Hartree-Fock) equations (e.g. [12]) where the effect of the motion of each quantum particle on the rest of the particles is accounted for in an averaged manner such that the detailed quantum correlations are ignored. One way to recover those is to introduce an additional degree of freedom such that besides the wave function each quantum particle is described also by a set of classical particles (walkers) which are guided by separate guiding wave functions in the spirit of de Broglie-Bohm wave mechanics. Then, if the i-th quantum particle is described by M walkers, we have for the walker's motion:

$$\mathbf{v}(\mathbf{r}_i^k) = \frac{\hbar}{m} \text{Im} \left[ \frac{1}{\varphi_i^k(\mathbf{r}_i,t)} \nabla_i \varphi_i^k(\mathbf{r}_i,t) \right]_{\mathbf{r}_i = \mathbf{r}_i^k(t)}, \tag{4}$$

where k=1,…,M denote the walkers assigned to the i-th quantum particle. Then the many-body wave function maps to a set of many-body wave functions for each walker (k) according to:

$$\Psi^k(\mathbf{r}_1, \mathbf{r}_2, ..., \mathbf{r}_N, t) = \prod_{i=1}^{N} \varphi_i^k(\mathbf{r}_i, t). \tag{5}$$

Therefore, we have multiple configurations of particles and guide waves to be propagated by appropriate evolutionary equations which in the TDQMC formalism consist of a set of (nonlinear) Schrödinger type of equations [10,11]:

$$i\hbar \frac{\partial}{\partial t} \varphi_i^k(\mathbf{r}_i,t) = \left[ -\frac{\hbar^2}{2M_i} \nabla_i^2 + \sum_{j \neq i}^{N} V^{eff}[\mathbf{r}_i, \mathbf{r}_j^k(t)] + V_{ext}(\mathbf{r}_i,t) \right] \varphi_i^k(\mathbf{r}_i,t), \tag{6}$$

where $V^{eff}[\mathbf{r}_i, \mathbf{r}_j^k(t)]$ is the effective potential experienced by the walkers from the *i*-th electron ensemble due to the walker's trajectories belonging to the rest of the electrons $\mathbf{r}_j^k(t)$, and $V_{ext}(\mathbf{r}_i,t)$ is a potential due to e.g. interaction with external electromagnetic fields. In this way the many-body Hamiltonian is reduced to one-body Hamiltonians where the degrees of freedom of the rest of the particles are expressed through their trajectories. The need for using effective potentials in Eq. (6) instead of true potentials is dictated by the quantum non-locality which is present in the Schrödinger equation, Eq. (1), through the dependence of the many-body wave function on the coordinates of all particles which makes it an object in configuration space. In TDQMC the effective potential is given by a Monte Carlo convolution of the true potential and a kernel function which incorporates the nonlocal quantum correlation length in a simple manner:

$$V^{eff}[\mathbf{r}_i, \mathbf{r}_j^k(t)] = (V * K)\left(\mathbf{r}_i, \mathbf{r}_j^k(t), \sigma_j^k\right), \tag{7}$$

where $V$ is the true interaction potential and $K$ is a normalized nonlocal kernel which depends on the nonlocal quantum correlation length $\sigma_j^k = \sigma_j^k(\mathbf{r}_j^k, t)$. In the Monte Carlo calculation the effective potential can be represented as a sum (Monte Carlo convolution) [11]:

$$V^{eff}[\mathbf{r}_i,\mathbf{r}_j^k(t)] = \frac{1}{Z_j^k}\sum_{l=1}^{M}V[\mathbf{r}_i,\mathbf{r}_j^l(t)]\,K\left[\mathbf{r}_j^l(t)-\mathbf{r}_j^k(t),\sigma_j^k\right], \quad (8)$$

where:

$$Z_j^k = \sum_{l=1}^{M}K\left[\mathbf{r}_j^l(t),\mathbf{r}_j^k(t),\sigma_j^k\right]$$

is the weighting factor. It is important to point out that the Monte Carlo convolution in Eq. (8) contains an implicit dependence on the probability density, which is easily seen when writing it as an integral:

$$V^{eff}[\mathbf{r}_i,\mathbf{r}_j^k(t)] = \int\left|\varphi_j(\mathbf{r},t)\right|^2 V[\mathbf{r}_i,\mathbf{r}]K\left[\mathbf{r}-\mathbf{r}_j^k(t),\sigma_j^k\right]d\mathbf{r}, \quad (9)$$

where $\left|\varphi_j(\mathbf{r},t)\right|^2$ denotes the joined space-time distribution of all trajectories $\mathbf{r}_j^l(t)$, $l=1,2,...,M$. This distinguishes the Monte Carlo convolution from the standard convolution used in e.g. the path integral formulation [13] where the probability density is not present under the integral in Eq. (9). It is assumed in TDQMC that the nonlocal quantum correlation length for the j-th species $\sigma_j^k(\mathbf{r}_j^k,t)$ can be characterized solely by the statistical properties of the distribution of that specie, e.g. by the kernel density estimation bandwidth or the standard deviation of the Monte Carlo sample:

$$\sigma_j^k(\mathbf{r}_j^k,t) = \alpha_j.\sigma_j(t), \quad (10)$$

where in general $\sigma_j^k(\mathbf{r}_j^k,t)$ should be a tensor with respect to space. Note that for the case where the kernel function tends to unity $K\left[\mathbf{r}-\mathbf{r}_j^k(t),\sigma_j^k\right]\to 1$ as $\sigma_j^k\to\infty$ (for e.g Gaussian kernel [11]), the effective potential in Eqs.(8),(9) reduces to the mean-field Hartree potential:

$$V^{eff}(\mathbf{r}_i) = \int\left|\varphi_j(\mathbf{r},t)\right|^2 V(\mathbf{r}_i,\mathbf{r})d\mathbf{r}, \quad (11)$$

in which case the guiding waves $\varphi_i^k(\mathbf{r}_i,t)$ coincide for all k. In the opposite case where $\sigma_j^k\to 0$ we have $K\left[\mathbf{r}-\mathbf{r}_j^k(t),\sigma_j\right]\to\delta\left(\mathbf{r}-\mathbf{r}_j^k(t)\right)$ which leads to the classical local potential:

$$V^{eff}\left(\mathbf{r}_i,\mathbf{r}_j^k(t)\right) = V\left[\mathbf{r}_i,\mathbf{r}_j^k(t)\right]. \quad (12)$$

In the latter case the equations (Eq.(6)) become decoupled and linear with respect to the guide waves. Equation (12) also implies that the Monte Carlo walkers which belong to different species experience a local pair-wise interaction which is closer to the classical case (ultra-correlated approximation). It is seen from Eqs.(9)-(12) that an opportunity is provided within TDQMC to smoothly tune the degree of correlation in the quantum system under consideration by changing the value of the nonlocal quantum correlation length $\sigma_j^k$. For each interaction potential there exists an optimal value of $\sigma_j^k$ which can be determined by variationally minimizing the ground state energy. For example, for an atom the average energy at moment $\tau$ can be estimated using:

$$E = \frac{1}{M} \sum_{k=1}^{M} \left[ \sum_{i=1}^{N} \left[ -\frac{\hbar^2}{2M_i} \frac{\nabla_i^2 \varphi_i^k(\mathbf{r}_i^k)}{\varphi_i^k(\mathbf{r}_i^k)} + V_{e-n}(\mathbf{r}_i^k) \right] + \sum_{i>j}^{N} V_{e-e}(\mathbf{r}_i^k - \mathbf{r}_j^k) \right]_{\substack{\mathbf{r}_i^k = \mathbf{r}_i^k(\tau) \\ \mathbf{r}_j^k = \mathbf{r}_j^k(\tau)}} \qquad (13)$$

where $V_{e-n}$ and $V_{e-e}$ are the electron-nuclear and electron-electron potentials, respectively.

Besides the quantum drift due to the guiding (Eq. (4)) the walkers are subject to quantum diffusion which is accounted for by adding a random component to walker's motion at each imaginary time step according to (see e.g. [14]):

$$d\mathbf{r}_i^k = \mathbf{v}(\mathbf{r}_i^k) dt + \mathbf{\eta} \sqrt{\frac{\hbar}{M_i} dt}, \qquad (14)$$

where Markovian process is assumed where $\mathbf{\eta}$ is a random variable with zero mean and unit variance. During the ground state preparation in TDQMC each walker samples the probability density given by the modulus square of the corresponding guide wave through the Metropolis algorithm. In this way in TDQMC the walker distribution may follow the true probability density unlike in other quantum Monte Carlo approaches where multi-dimensional trial wave functions are employed which can provide accurate energies for not that accurate walker's distributions.

It is clear from Eq. (14) that as $\hbar/M_i \to 0$ the quantum diffusion in Eq. (6) decreases and so does the width of the corresponding distribution $\left|\varphi_i^k(\mathbf{r}_i,t)\right|^2$. Therefore in this case the standard deviation $\sigma_j$ vanishes together with the nonlocal quantum correlation length $\sigma_j^k$, and as a result, the ensemble of walkers for each species collapses to a single walker thus transforming the quantum ensemble to an ensemble of classical particles with the only force between these being due to the classical potential $V(\mathbf{r}_i, \mathbf{r}_j)$, (see Eq. (12)).

### 3. Finite temperature formulation

In order to account for the finite temperature effects within the TDQMC methodology we can immerse the quantum system under consideration in a thermal environment which consists of bath of harmonic oscillators [5]. The bath can be considered classically or quantum mechanically, depending on the way the oscillators are treated. In the more rigorous quantum approach, the bath obeys a set Schrödinger equations coupled to the equations of the system through the assumption that the system and the bath comprise a closed system with a common wave function which is next expanded as in Eq. (5) but with additional degrees to account for the bath variables [15]:

$$\Psi^k(\mathbf{r}_1, \mathbf{r}_2, ..., \mathbf{r}_N, \mathbf{R}_1, ..., \mathbf{R}_L, t) = \prod_{i=1}^{N} \varphi_i^k(\mathbf{r}_i, t) \prod_{j=1}^{L} \Phi_j^k(\mathbf{R}_j, t), \qquad (15)$$

where, in the spirit of TDQMC, the new functions $\Phi_j^k(\mathbf{R}_j, t)$ represent the guide waves for the bath walkers, for $L$ bath oscillators in Eq. (15). Then, the set of equations (6) should be expanded by the addition of $L$ new Schrödinger equations for the bath wave functions $\Phi_j^k(\mathbf{R}_j, t)$ to be solved self consistently with the system equations. An advantage of that approach is that at the zero temperature limit the non-trivial zero fluctuations of the quantum bath can be accounted for. In the product expansion of Eq. (15) it is implicitly assumed that the $k$-th walker from the $i$-th specie interacts mostly with the $k$-th walker of the rest of species including also the bath oscillators. Of course some other walkers around the $k$-th walker are always included in the interaction due to the quantum non-locality described by the non-local quantum correlation length as discussed above. On the other hand it has been shown that within the mean field approximation the system-bath interaction can be reduced to Langevin-Schrödinger equation where the bath-induced noise dominates over the quantum system-bath correlations [15, 16].

Since our aim here is to establish the frame of feasibility of TDQMC for non-zero temperature we focus first on the finite temperature ground state preparation. We use a generic system plus bath Hamiltonian with bilinear coupling between the system species and the bath oscillators, which has proved useful in various fields:

$$H = \sum_{i=1}^{N}\left[-\frac{\hbar^2}{2m_i}\nabla_i^2 + V_S(\mathbf{r}_i)\right] + \sum_{j=1}^{L}-\frac{\hbar^2}{2M_j}\nabla_j^2 + H_{SB}(\mathbf{r}_i,\mathbf{R}_j) + V_{ext}(\mathbf{r}_i,t), \quad (16)$$

where for bath oscillators with mass $M_j$ and frequency $\Omega_j$ the system-bath Hamiltonian reads [5,17]:

$$H_{SB}(\mathbf{r}_i,\mathbf{R}_j) = \sum_{j=1}^{L}V_{SB}(\mathbf{r}_i,\mathbf{R}_j) = \sum_{j=1}^{L}\frac{M_j\Omega_j^2}{2}\left(\mathbf{R}_j - \frac{C_{ij}\mathbf{r}_i}{M_j\Omega_j^2}\right)^2, \quad (17)$$

where $C_{ij}$ are the coupling strength constants.

In accordance with the product expansion of Eq. (15) there are two groups of TDQMC equations for the guide waves, one group of $N$ Schrödinger equations for the system species with masses $m_i$:

$$i\hbar\frac{\partial}{\partial t}\varphi_i^k(\mathbf{r}_i,t) = \left[-\frac{\hbar^2}{2m_i}\nabla_i^2 + V_S(\mathbf{r}_i) + \sum_{j\neq i}^{N}V_{SS}^{eff}[\mathbf{r}_i,\mathbf{r}_j^k(t)] + \sum_{j=1}^{L}V_{SB}[\mathbf{r}_i,\mathbf{R}_j^k(t)]\right]\varphi_i^k(\mathbf{r}_i,t) \quad (18)$$

and another group of $L$ Schrödinger equations for the guide waves of the bath oscillators:

$$i\hbar\frac{\partial}{\partial t}\Phi_j^k(\mathbf{R}_j,t) = \left[-\frac{\hbar^2}{2M_j}\nabla_j^2 + \sum_{l=1}^{N}V_{SB}[\mathbf{R}_j,\mathbf{r}_l^k(t)]\right]\Phi_j^k(\mathbf{R}_j,t), \quad (19)$$

where we have excluded in Eq. (19) the direct interaction between the bath oscillators. Equations (18) and (19) are then solved self-consistently with the guiding equations for the velocities of the system walkers:

$$\mathbf{v}(\mathbf{r}_i^k) = \frac{\hbar}{m}\text{Im}\left[\frac{1}{\varphi_i^k(\mathbf{r}_i,t)}\nabla_i\varphi_i^k(\mathbf{r}_i,t)\right]_{\mathbf{r}_i=\mathbf{r}_i^k(t)}, \quad (20)$$

and for the bath walkers:

$$\mathbf{V}(\mathbf{r}_j^k) = \frac{\hbar}{M_j}\text{Im}\left[\frac{1}{\Phi_j^k(\mathbf{R}_j,t)}\nabla_j\Phi_j^k(\mathbf{R}_j,t)\right]_{\mathbf{R}_j=\mathbf{R}_j^k(t)} \quad (21)$$

One possible strategy to deal with the system-bath interaction would be to try to eliminate the bath, which would result in a set of coupled Langevin-Schrödinger equations for the system alone [15,16]. Instead, we choose here to integrate the whole system of equations (18)-(21), after posing some physical prerequisites on the role of the system-bath interaction. While in Eq. (18) and (19) the nonlocal effective potential $V_{SS}^{eff}$ between the system species is determined according to the general rules discussed in the Introduction, we have some freedom about the choice of the system-bath effective potential $V_{SB}^{eff}$ which determines the degree of system-bath correlations. In case we are not interested in the detailed system-bath correlations $V_{SB}^{eff}$ reduces to the mean-field Hartree potential which would mean that all guide waves for the bath $\Phi_j^k(\mathbf{R}_j,t)$ coincide for all $k$, for each given oscillator $j$:

$$i\hbar\frac{\partial}{\partial t}\varphi_i^k(\mathbf{r}_i,t) = \left[-\frac{\hbar^2}{2m_i}\nabla_i^2 + V_S(\mathbf{r}_i) + \sum_{j\neq i}^{N} V_{SS}^{eff}[\mathbf{r}_i,\mathbf{r}_j^k(t)] + \frac{1}{M}\sum_{l=1}^{M} V_{SB}[\mathbf{r}_i,\mathbf{R}_i^l(t)]\right]\varphi_i^k(\mathbf{r}_i,t) \quad (22)$$

$$i\hbar\frac{\partial}{\partial t}\Phi_j(\mathbf{R}_j,t) = \left[-\frac{\hbar^2}{2M_j}\nabla_j^2 + \frac{1}{M}\sum_{l=1}^{M} V_{SB}[\mathbf{R}_j,\mathbf{r}_j^l(t)]\right]\Phi_j(\mathbf{R}_j,t) \quad (23)$$

The equations for the bath degrees can be simplified further for a classical bath where, for $\hbar/M_j \to 0$, we arrive at:

$$M_j\ddot{\mathbf{R}}_j = -\nabla_j\left[\frac{1}{M}\sum_{l=1}^{M} V_{SB}\left[\mathbf{R}_j,\mathbf{r}_j^l(t)\right]\right]_{\mathbf{R}_j=\mathbf{R}_j(t)} \quad (24)$$

Oppositely, in this work we shall assume that each walker for a given system specie is connected to only one oscillator from the bath, which yields from Eqs.(18), (19):

$$i\hbar\frac{\partial}{\partial t}\varphi_i^k(\mathbf{r}_i,t) = \left[-\frac{\hbar^2}{2m_i}\nabla_i^2 + V_S(\mathbf{r}_i) + \sum_{j\neq i}^{N} V_{SS}^{eff}[\mathbf{r}_i,\mathbf{r}_j^k(t)] + V_{SB}[\mathbf{r}_i,\mathbf{R}_i^k(t)]\right]\varphi_i^k(\mathbf{r}_i,t) \quad (25)$$

$$i\hbar\frac{\partial}{\partial t}\Phi_j^k(\mathbf{R}_j,t) = \left[-\frac{\hbar^2}{2M_j}\nabla_j^2 + V_{SB}[\mathbf{R}_j,\mathbf{r}_j^k(t)]\right]\Phi_j^k(\mathbf{R}_j,t) \quad , \quad (26)$$

which yields for a classical bath ($\hbar/M_j \to 0$):

$$M_j\ddot{\mathbf{R}}_j^k = -\nabla_j\left[V_{SB}\left[\mathbf{R}_j^k,\mathbf{r}_j^k(t)\right]\right]_{\mathbf{R}_j^k=\mathbf{R}_j^k(t)} \quad (27)$$

It can be seen form Eqs.(23),(24) that for the mean-field bath in the classical case each oscillator can be represented by a single walker which experiences the averaged potential due to all system walkers (no upper index $k$ in Eq. (24)). The latter contrasts the case of Eq. (27) where there is still a pairwise classical interaction between the system and the bath walkers.

In general, the TDQMC methodology provides for each separate quantum particle a large ensemble of guide waves with statistical meaning as replicas of the true wave function. For an isolated particle all guide waves coincide because these describe a pure state and there is no need for density matrices while for an interacting particle in a mixed state the density matrix can easily be constructed from the guide waves without first calculating the density matrix of the whole system (by e.g. solving the corresponding master equation) and then tracing out the rest of the degrees. Since in the mixture of nonseparable system-bath states the probability of finding a given guide wave is determined by the classical probability which follows the probability distribution of the system walkers in space, the density matrix in coordinate representation for the $i$-th specie can be easily calculated [18]:

$$\rho_i(\mathbf{r}_i, \mathbf{r}'_i, t) = \frac{1}{M} \sum_{k=1}^{M} \varphi_i^{k*}(\mathbf{r}_i, t) \varphi_i^{k}(\mathbf{r}'_i, t) \tag{28}$$

In this way the calculation of the density matrix at each time step from Eq. (25) and Eq. (28) can provide vital information on the temperature dependence of the probability distribution in space as well as on the processes of decoherence and space-time correlations for the different quantum species. For finite temperature, the average energy of the system at ground state can be calculated using $\langle E \rangle = Tr(\rho H_S)$ which after substitution of Eq. (28) for the density matrix, and provided $\varphi_i^k(\mathbf{r}_i, t)$ are eigen-functions of the system Hamiltonian with eigen-energies $E_k$, yields for the average energy $\langle E \rangle = 1/M \sum_k E_k$, which in fact is equivalent to Eq. (13).

## 4. Numerical results

Known methods to access the temperature dependences in quantum systems include analytical solution of the Bloch equation for the density matrix for harmonic oscillator [19] and the quasi-classical expansions due to Feynman and Kleinert [20] and others [21], as well as other path integral techniques. Here we compare the TDQMC predictions for the temperature dependence of the diagonal elements of the system density matrix with the numerically exact results given by (hereafter we assume that all particles are electrons and use atomic units $e = \hbar = k_b = 1; m_i = M_j = 1$) [19]:

$$\rho(x,x,\beta) = Z^{-1} \sum_n |\psi_n(x)|^2 \exp(-\beta E_n), \tag{29}$$

where $\beta = 1/T$, $Z$ is the statistical sum, and $E_n$, $\psi_n(x)$ are calculated by numerically diagonalizing the system Hamiltonian. This allows one also to monitor the average energy at finite temperature $\langle E \rangle = 1/Q \sum_n E_n \exp(-\beta E_n)$ where $Q$ is the statistical sum.

In order to solve Eqs.(17), (25) and (26) self-consistently together with the guiding equations (20) and (21) in one spatial dimension we assume for simplicity that the bath oscillators are initially in coherent states:

$$\Phi_j(X_j,0) \sim \exp\left[-\frac{\Omega_j(X_j - \bar{X}_j)^2}{2} + i\bar{P}_j X_j\right], \tag{30}$$

where $X_j$ replaces $\mathbf{R}_j$, and $\bar{X}_j, \bar{P}_j$ are the average initial coordinate and momentum. We assume the bath frequencies $\Omega_j$ are equidistant (Ohmic bath [5]) and they span the available bound energy levels of the system while at the same time these are off-resonance with the levels. In this way through the joined solution of equations (17), (25) and (26) the back reaction of the bath oscillators causes dissipation and dephasing on the system electron in accordance with the fluctuation-dissipation theorem. Also, in some cases it is more convenient to simplify the interaction Hamiltonian in Eq. (17) to its more explicit bilinear form:

$$H_{SB}(\mathbf{r}_i, \mathbf{R}_j) = \sum_{j=1}^{L} \left[\frac{M_j \Omega_j^2}{2} \mathbf{R}_j^2 + C_{ij} \mathbf{r}_i \cdot \mathbf{R}_j\right], \tag{31}$$

which we employ in the present work.

By assigning a set of distinguishable wave functions to each interacting electron it is presumed by definition in TDQMC that the electron is generally in a mixed state described by a density matrix [18]. In case of system-bath interaction that mixed state can be easily found by attaching different bath oscillators to the different walkers and guide waves of the system, as described in Section 3. First we consider a model atom consisting of a single electron connected to massive core through a smoothed Coulomb potential in one spatial dimension [22]:

$$V_S(x) = -\frac{1}{\sqrt{1+x^2}},  \qquad (32)$$

where the system possesses bound spectrum with energy levels -0.669 a.u., -0.275 a.u., -0.147 a.u., -0.0583 a.u. Therefore in this paper we choose for the maximum bath frequency $\Omega_{j,\max} = 0.6 a.u.$ which ensures that none of the bath oscillators may cause direct transitions from the ground state to the continuum states of the model atom.

For finite temperature the joined system-bath ground state is prepared using imaginary-time evolution of an initial set of guide waves for both the system and the bath according to Eqs.(17), (25), (26) while each Monte-Carlo walker experiences quantum diffusion by sampling its own probability density distribution given by the modulus squared of the corresponding guide wave. Since for the inner atomic shells the relevant temperatures where higher energy levels can be populated are in the $10^4$ K range here we consider temperature interval $0 \leq T \leq 0.1$ a.u.

In order to show the nature of the mixed state in the presence of bath, we depict in Fig. 1 five randomly chosen probability densities $\left|\varphi_1^k(x_1,t)\right|^2$; k=1..5, after the finite temperature ground state of the atom has been established, for $\beta = 10 a.u.$. It is seen that these densities differ significantly (as expected for a mixed state) due to the different system-bath potentials $V_{SB}$ experienced by the guide waves in the TDQMC equations, Eqs. (25), (26). Since the density matrix (Eq. (28)) reflects the statistical properties of the wave ensemble, the distribution of waves in Fig.1 will generally broaden the overall probability density in space given by the diagonal elements $\rho(x,x,\beta)$. In Fig.2a we have plotted the distribution $\rho(x,x,\beta)$ for three temperatures according to the numerical solution of the TDQMC equations as compared to the numerically exact results provided by Eq. (29) where the system Hamiltonian is being diagonalized using an implicitly restarted Arnoldi method (package ARPACK [23]). The exact results are drawn with black lines to be compared with the TDQMC results (red lines). It is noteworthy that for lower temperature ($\beta = 100 a.u.$) all guide waves for the system converge to a single wave, thus proving the ease with which

TDQMC treats low temperatures, unlike when using path integral formulation which may become unsatisfactory for $\beta > 15 a.u.$ [24]. It is seen form Fig.2a that for the other two cases $\beta = 11.7 a.u.$ and $\beta = 10 a.u.$ there is also a very good correspondence between the TDQMC predictions and the exact results. It is important to stress that just matching the peak values of the TDQMC distributions with the exact ones by varying the coupling constants $C_{ij}$ would not be sufficient because the shapes and in particular the widths of those is what matters when considering mixed states where many wave functions contribute to the density matrix. Figure 2b shows the distributions $\rho(x, x, \beta)$ for low temperature ($\beta = 100 a.u.$) and for $\beta = 10 a.u.$, for two electrons in the same core potential (Eq. (31)) but also interacting with each other through a smoothed Coulomb potential:

$$V_{SS}[x_1, x_2^k(t)] = \frac{0.2}{\sqrt{1 + \left[x_1 - x_2^k(t)\right]^2}}. \tag{33}$$

It is seen from Fig.2b that although the ground state distributions for the two different temperatures are broader and lower than in Fig.2a, which is attributed to the electron-electron repulsion, the TDQMC predictions are again in a very good agreement with the exact results. The above proof-of-principle examples show that the TDQMC method is capable of providing rather correct results for the ground state of systems of interacting quantum particles at finite temperature.

In order to demonstrate the dynamic behavior of one or two quantum particles at finite temperature we excite the atom with a very short uni-polar (no carrier) electromagnetic pulse and next observe the oscillations of the induced dipole moment calculated using the definition $\langle x \rangle = Tr(x\rho)$. The blue lines in Fig.3a,b show the dipole oscillations for a single electron and for two interacting electrons, for zero bath temperature. It is seen that these oscillations practically do not fade out within a few-femtosecond time scale, which can be expected for almost pure quantum states where there is a little dephasing due to the zero fluctuations of the bath oscillators. This is in contrast to the finite temperature cases (T=0.1 a.u. in Fig.3a and T=0.05 a.u. in Fig.3b) where there is a visible relaxation of the dipole moment oscillations (red lines) which is due to the mutual dephasing within the ensemble of waves which constitute the mixed state (see Fig.1), which is attributed to the rapid dissipation in the whole system in accordance with the fluctuation-dissipation theorem.

## 5. Conclusions

In this work the time-dependent quantum Monte Carlo method was adopted to quantum systems at finite temperature using a bath of quantum harmonic oscillators. Our results show that TDQMC can robustly accommodate the large number of degrees for the bath oscillators (~ 10 000) together with the quantum system, in a self-consistent manner. The mixed quantum states which arise due to the system-bath interaction appear as a natural result within the TDQMC framework where the corresponding single-particle density matrices are easily calculated. The obtained test results are encouraging in that they predict correctly the broadening of the particle distributions for finite temperatures as well as the decay of the oscillations of the dipole moment as a result of the decoherence in the guide-wave ensembles due to both the system-bath interaction and the interactions within the system itself. The approach reported here can be applied to study transient phenomena in presence of real phonons in condensed phases as well as in external time-dependent fields.

## 6. Acknowledgment

The author gratefully acknowledges support from the Bulgarian National Science Fund under grant FNI T02/10.

# References

[1] H. Haug, A. P. Jauho, Quantum Kinetics in Transport and Optics of Semiconductors, Springer, Berlin, 1996.

[2] R. M. Fye, J. E. Hirsch, Phys. Rev. B 38 (1988) 433.

[3] P. Werner, A. J. Millis, Phys. Rev. Lett. 99 (2007) 146404.

[4] M. D. Kostin, J. Chem. Phys. 57 (1972) 3589.

[5] U. Weiss, Quantum Dissipative Systems, in: Series in Modern Condensed Matter Physics, vol. 13, World Scientific, Singapore, 2008.

[6] M. Nest, H.-D. Meyer, J. Chem. Phys. 119 (2003) 24.

[7] L. Muhlbacher, E. Rabani, Phys. Rev. Lett. 100 (2008) 176403.

[8] N. Makri, Comp. Phys. Commun. 63 (1991) 389.

[9] I. P. Christov, Opt. Express 14 (2006) 6906.

[10] I. P. Christov, J. Chem. Phys. 128 (2008) 244106.

[11] I. P. Christov, J. Chem. Phys. 136 (2012) 034116.

[12] B. H. Bransden and C. J. Joachain, Quantum Mechanics, 2nd edition, Pearson Education Limited, London, 2000.

[13] R. P. Feynman and A. R. Hibbs, Quantum Mechanics and Path Integrals, McGraw-Hill, New York, 1965.

[14] B. Hammond, W. Lester, and P. Reynolds, Monte Carlo Methods in Ab Initio Quantum Chemistry, World Scientific, Singapore, 1994.

[15] W. H. Miller, in Stochasticity and Intramolecular Redistribution of Energy, edited by R. Lefebvre and S. Mukamel, Reidel, Dordtrecht, 1987.

[16] U. Peskin, M. Steinberg, J. Chem. Phys. 109 (1998) 704.

[17] A. O. Caldeira, A. J. Leggett, Physica A 121 (1983) 587.

[18] I. P. Christov, Phys. Scr. 91 (2016) 015402.

[19] R. P. Feynman, Statistical Mechanics, Benjamin, Massachusetts, 1972.

[20] R. P. Feynman, H. Kleinert, Phys. Rev. A 34 (1986) 5080.

[21] A. Cuccoli, A. Macchi, V. Tognetti, M. Neumann, R. Vaia, Phys. Rev. B 45 (1992) 2088.

[22] R. Grobe, J. H. Eberly, Phys. Rev. Lett. 68 (1992) 2905.

[23] R. B. Lehoucq, D. C. Sorensen, C. Yang, ARPACK Users' Guide, SIAM, Philadelphia, 1998.

[24] W. Janke and H. Kleinert, Phys. Lett. 118 (1986) 371.

**Figure captions:**

**Figure 1**. Probability densities of five randomly chosen guide waves for an electron at ground state at temperature T=0.1 a.u. (3.158 $10^4$ K).

**Figure 2.** Probability densities for a single electron at ground state for different temperatures -(a), and for an electron in a two-electron atom -(b). Black lines -exact result, red lines -from TDQMC density matrices. The insets show the whole distributions while the main plots show their tops, for better comparison. The total number of bath oscillators in these calculations is typically ~ 10 000.

**Figure 3**. Oscillations of the dipole moment of the electron induced by a very short uni-polar electromagnetic pulse to shift the electron from its equilibrium position, for a single-electron atom -(a), and for two-electron atom -(b). Blue lines -for zero-temperature bath, red lines –for T=1 a.u. -(a), and for T=0.5 a.u. -(b).

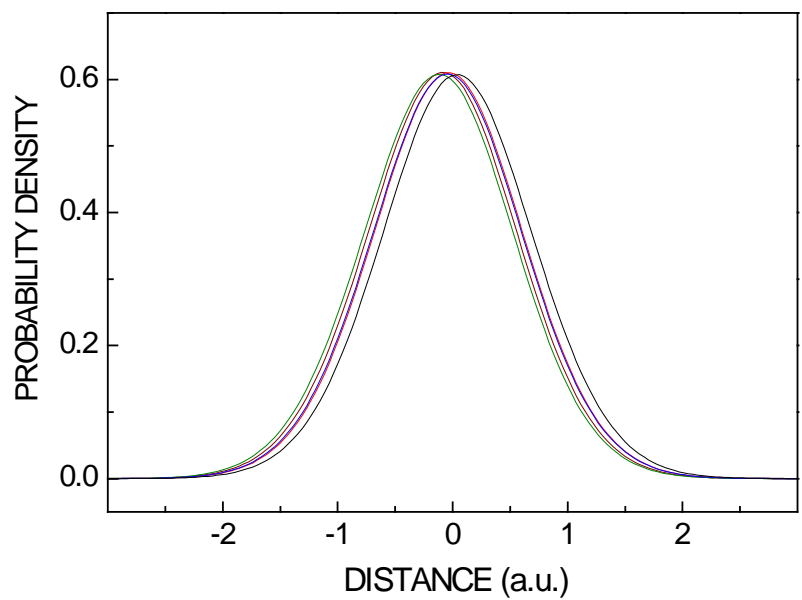

Fig. 1

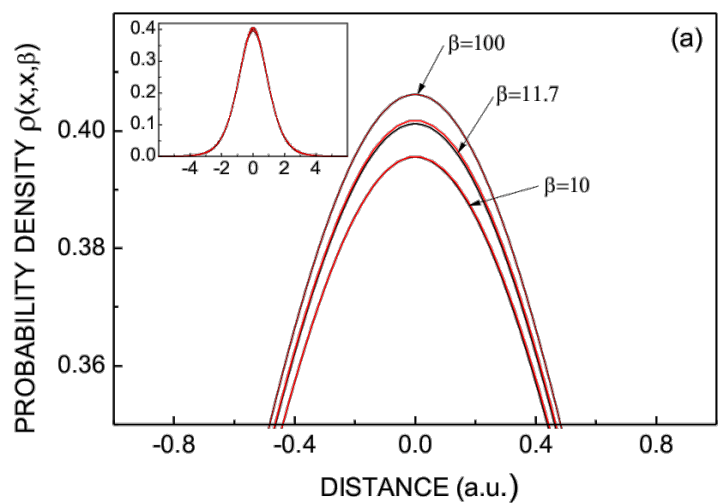
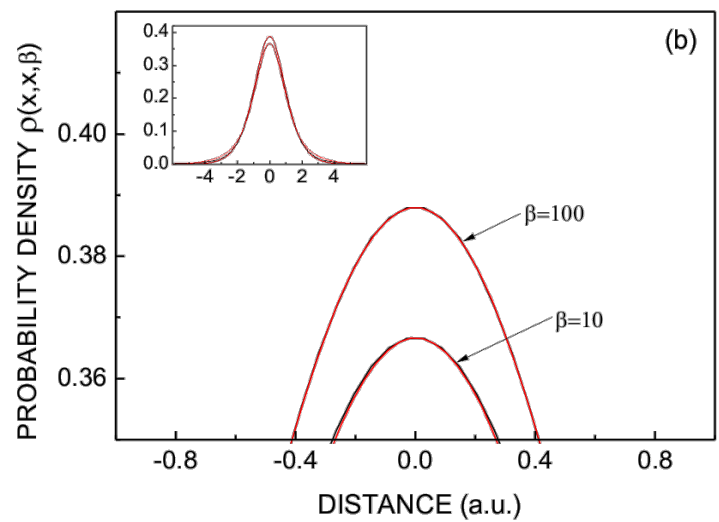

Fig. 2

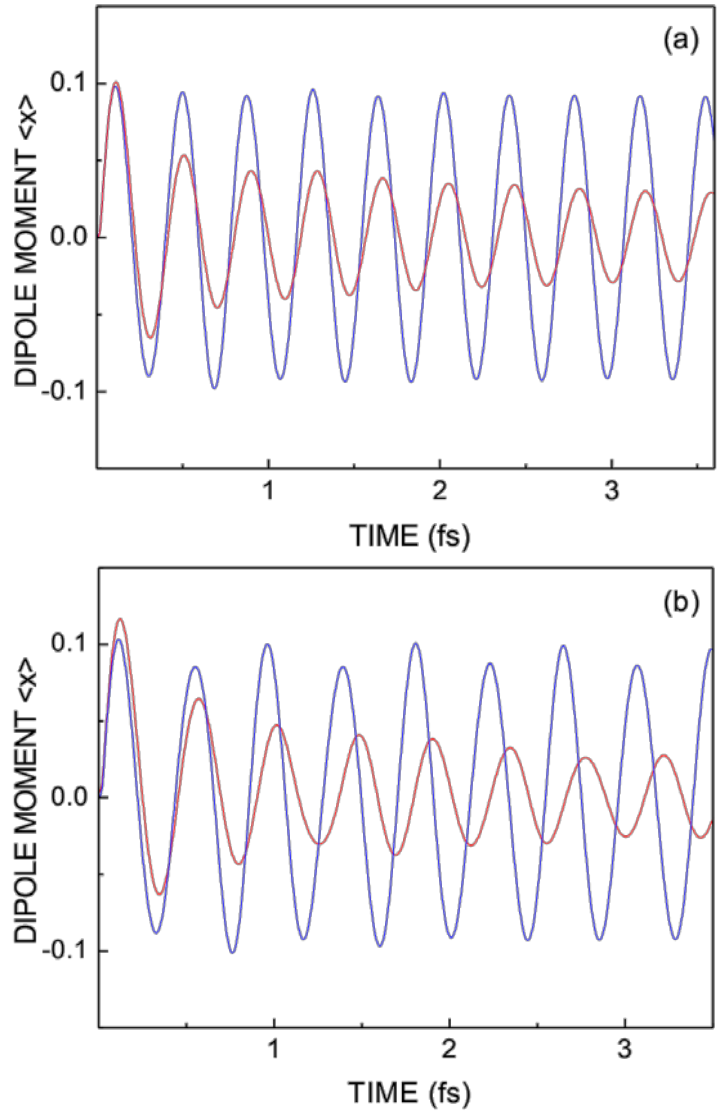

Fig. 3